\documentclass[10pt,romanappendices,conference,twocolumn]{IEEEtran}
\usepackage{cite}
\usepackage{amssymb}
\usepackage{graphicx}
\usepackage{amssymb}
\usepackage{amsbsy}
\usepackage{float}
\usepackage{balance}
\usepackage{subfigure}
\usepackage{url}
\usepackage{bm}
\newfloat{longequation}{b}{ext}
\newfloat{longequation}{t}{ext}

\usepackage[cmex10]{amsmath}

\usepackage{array}
\usepackage{varioref}
\usepackage[normalem]{ulem}
\graphicspath{{./figs/}}
\usepackage{algorithm2e}

\begin{document}

\title{User Scheduling \\ for Coordinated   Dual Satellite Systems\\with Linear Precoding }

\author{\IEEEauthorblockN{ Dimitrios Christopoulos\IEEEauthorrefmark{1}, Symeon Chatzinotas\IEEEauthorrefmark{1} and Bj\"{o}rn Ottersten\IEEEauthorrefmark{1}\IEEEauthorrefmark{3}}
\IEEEauthorblockA{\IEEEauthorrefmark{1}SnT - securityandtrust.lu,  University of Luxembourg
\\email: \textbraceleft dimitrios.christopoulos, symeon.chatzinotas, bjorn.ottersten\textbraceright@uni.lu}
\IEEEauthorblockA{\IEEEauthorrefmark{3}Royal Institute of Technology (KTH), Sweden,
email: bjorn.ottersten@ee.kth.se}
}
\maketitle


\begin{abstract}
The constantly increasing demand for interactive broadband satellite communications is driving current research to explore novel system architectures that  reuse frequency in a more aggressive manner. To this end, the topic of dual satellite systems, in which satellites share spatial (i.e. same coverage area) and spectral (i.e. full frequency reuse) degrees of freedom is introduced.   In each multibeam satellite,    multiuser interferences are mitigated by employing zero forcing precoding with realistic per antenna power constraints. However,  the two sets of users that the transmitters are separately serving, interfere. The present contribution, proposes  the partial cooperation, namely coordination between the two coexisting transmitters in order to reduce interferences and enhance the performance of the whole system, while maintaining moderate system complexity. In this direction, a heuristic, iterative, low complexity algorithm that allocates users in the two interfering sets is proposed.  This novel algorithm,  improves the performance of each satellite and of  the overall system, simultaneously. The first is achieved by maximizing the orthogonality between users allocated in the same set, hence optimizing the zero forcing performance,  whilst the second by minimizing the level of interferences between the two sets. Simulation results show that the proposed method, compared to conventional techniques, significantly increases spectral efficiency.
\end{abstract}
\section{Introduction}

Towards the next generation of broadband multibeam Satellite Communication  (SatCom) systems, innovative system architectures need to be considered in order to meet the highly increasing demand for throughput and close the digital divide. Spectrum scarcity is a major obstacle, especially in a satellite context where the higher frequency bands exhibit challenging channel impairments. In this direction, the investigation of full frequency reuse techniques that exploit the spatial degrees of freedom offered by the multibeam antenna is necessitated. Furthermore, in the evolution of geostationary (GEO) satellite systems, orbital slot congestion is an uprising problem. The fact that unpredictable changes in the traffic demand might cause the launch of secondary satellites that support existing ones should be considered as well. More importantly, long periods of coexisting satellites appear during the replacement phase of old satellites. Finally,  cooperation between multibeam satellites has the potential of overcoming the major issue of Adjacent Satellite Interferences (ASI). It should also be noted that current research in cognitive satellite communications is considering scenarios where two satellites coexist \cite{Sharma2012}. All the above arguments
provide reasonable cause towards the investigation of the optimal coexistence of two satellites.

Existing
literature on Multiuser Multiple Input Multiple Output (MU-MIMO) antennas, provides transmitter techniques to alleviate the multiuser interferences in many cases \cite{Costa1983}.  Furthermore, linear low complexity techniques have proven more realistic in terms of practical  implementation. Specifically, focusing on the forward link (FL) of multibeam SatComs, linear joint processing techniques have shown great potential by providing a substantial tradeoff between implementation complexity and near optimal performance in terms of Sum-Rate (SR) \cite{Christopoulos2012}. Specifically, Zero Forcing (ZF) precoding, in detail analyzed further on,  is based on channel inversion and then optimal power allocation over the users, with the aim of maximizing some performance metric. The metrics commonly addressed in  literature involve either the total throughput performance (i.e. max SR criterion)  or the Signal to Interference plus Noise Ratio (\( \mathrm{SINR}\)) level of the worst   user (i.e. max fairness criterion). Another important parameter in linear precoding is the type of constraints that will be assumed. Usually, a total sum power constraint simplifies the analysis and provides better results since the available power is freely allocated in every antenna. Despite its performance, the sum power constraint is unrealistic since each satellite antenna is fed by a dedicated high power amplifier (HPA) operating close to saturation. Hence, power cannot be allocated freely amongst the transmit antennas and a per antenna power constraint should be considered.
In the present work, each satellite employs ZF beamforming  while the power allocation is optimized under max SR criteria, subject to realistic per antenna constraints.

Albeit the aforementioned technique, in a dual satellite system, intersatellite interferences still need to be handled. To completely mitigate these interferences, without  frequency orthogonalization,  full cooperation between the two transmitters has to be employed. Subsequently, the cooperating multi-antenna transmitters should perform joint or coherent transmission
to all users while,   both data and channel state information (CSI) must be exchanged\cite{Huang2011}.  In a SatCom context, each satellite is served by one or multiple dedicated gateways, thus  a fully cooperative dual satellite system would require a large number of interconnected GWs that exchange a substantial load of information.  In the light of the above observations, partial cooperation (i.e. coordination) between the two satellites is proposed, therefore  reducing the amount of data exchanged.

The rest of the present paper is structured as follows. A brief review of the existing related work is provided in Section \ref{sec: Related work}. The considered system model is described with detail in Section \ref{sec: System model}. Section \ref{sec: SIUA} explores the theoretical aspects of the developed heuristic algorithm, explaining the intuition behind it.  In Section \ref{sec: Simulation Results}, simulation results for the performance of the proposed algorithm are presented. Conclusions are drawn in Section \ref{sec: conclusions}, along with future extensions of the present work.

\section{Related Work}\label{sec: Related work}
Linear precoding techniques and especially ZF have
been extensively investigated for terrestrial systems in \cite{Wiesel-08,Yoo2006b} and the references therein.
Despite their suboptimality, these techniques can still achieve asymptotically optimal performance under specific conditions, as proven in \cite{Yoo2005,Yoo2006b,Ng_TIT_08}. Specifically, if user channels are perfectly orthogonal to each other, ZF will attain maximum performance. Under the assumption of large random user sets, the probability of orthogonal users increases. User selection, however, is a highly complex problem. Despite its complexity, simple suboptimal  algorithms in the existing literature provide substantial gains with affordable complexity. Based on existing algorithms, \textit{Yoo et al} \cite{Yoo2005} proposed an iterative user selection algorithm that allows ZF to achieve the performance of non-linear precoding\cite{Weingarten2006a}  when the number of available for selection users grows to infinity.


Despite the extensive literature on linear precoding and user selection, herein we further consider the optimal allocation of the selected users in two coexisting groups that interfere with each other. The novelty of the proposed work lies in the fact that user selection and allocation, not only optimizes the ZF performance of each system but also considers the interaction between the two transmitters, i.e. the inter-satellite interference. The procedure of selecting users out of a large pool and allocating them to specific sets is referred to as user scheduling.



\textit{Notation}: Throughout the paper, \(\mathcal{E}[\cdot]\),  \(\left(\cdot\right)^\dag\), \(||\cdot||\)   denote the expectation, the conjugate transpose and the norm operations, respectively.
Bold face lower case characters denote column vectors and upper case denote matrices while 
$\mathbf{I}_n$ denotes an identity matrix of size $n$. Upper case calligraphic characters denote sets. Operations $\mathcal A=\emptyset\ $, and $\mathcal {A-B}$ define an empty set and the relative complement of $\mathcal B$ in $\mathcal A$, respectively. Finally  $|\mathcal A|$, denotes the cardinality of a set.
\section{System Model}\label{sec: System model}
\begin{figure}
  \centering
  \includegraphics[width=1\columnwidth]{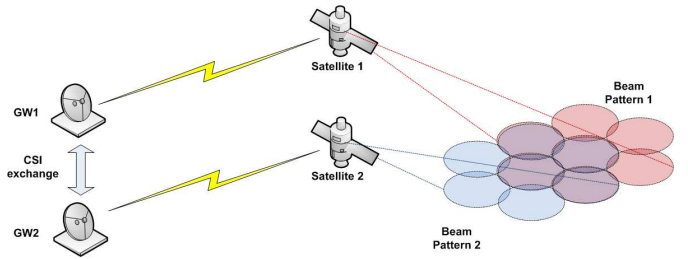}\\
  \caption{System model}
\label{fig: dual satellite}
\end{figure}

The system under investigation consists of two collocated multibeam satellites  with overlapping coverage areas,  serving  fixed single antenna users (Fixed Satellite Services, FSS). A large number of users uniformly distributed in each beam is assumed and a Time Division Multiple Access (TDMA) scheme is realized, leading to one user per beam served, during each time slot.  An overview of the considered system is depicted in Fig. \ref{fig: dual satellite}, where the focus is on the Forward Link (FL) downlink of the satellites (i.e. the link between the satellite and the users), while the FL uplink, or  feeder link (i.e. the link between each satellite and the   earth gateway station), is considered ideal.
It should be clarified that the present work for simplicity purposes, does not model the earth curvature and the satellite orbit geometry. Subsequently, the variations in the distances of the beam centers and the distance between the satellites are not modeled but will be handled in future extensions.

Considering  each multibeam satellite separately, linear precoding is employed to cancel out multiuser interferences.  By denoting
$N \ \text{and} \ K$ the number of transmit antennas and single antenna users respectively, the one user per beam per timeslot assumption implies a symmetric system, i.e. $N = K$. Subsequently, in each separate transmitter, a Multi-User (MU) Multiple Input Single Output (MISO) Broadcast Channel (BC) is realized and   the input-output analytical expression for the   \(k\)-th user reads as
\begin{equation}\label{eq: General input-output}
y_{k}= \mathbf h^{\dag}_{k}\mathbf x+n_{k},
\end{equation}
where \(\mathbf h^{\dag}_{k}\) is an \(1 \times N\) vector composed of the channel coefficients between the \(k\)-th user and the  \(N\) antennas (i.e. feeds) of the satellite, \(\mathbf x\) is an \(N \times 1\)  vector of transmitted symbols and  \(n_{k}\) is the independent identically distributed (i.i.d) zero mean  Additive White Gaussian Noise (AWGN)  measured at the \(k\)-th user's receive antenna. The noise is assumed normalized, thus $ \mathcal E\left\{|n_k|^2\right\}=1 $.

To accurately model the multibeam satellite channel the following considerations are made. Under the assumption of fixed users with highly directive antennas, the satellite channel can be modeled  as AWGN channel with real channel  gains\footnote{The main conclusions of this study, still apply if a random phase is incorporated in the channel gains.} that depend only on the multibeam antenna pattern and on the user position.  The elements of the $k$-th user's channel vector are  the square roots of the gain coefficients calculated using the well accepted method of Bessel functions \cite{Diaz2007}
\begin{equation}\label{eq: beam_gain}
g_{ik}(\theta_{ik})=G_{\max}\left(\frac{J_1(u)}{2u}+36\frac{J_3(u)}{u^3}\right)^2,
\end{equation}
where \(u=2.07123\sin\theta/\sin\theta_{3dB}\), \(J_1\), \(J_3\) are the Bessel functions of the first kind, of order one and three respectively and \(G_{\max}\) is the maximum axis gain of each antenna. The $k$-th users' position  corresponds to an off-axis angle \(\theta_{ik}\)   with respect to the boresight of the $i$-th beam where $\theta_{i}=0^{\circ}$.

\subsection{Transmit beamforming}
Transmit beamforming is a multiuser precoding technique that separates user data streams in different beamforming directions\cite{Yoo2006b}. Let us denote $s_k$, $\mathbf w_k$ and $p_k$ as the unit power symbol, the $N \times 1 $ transmit precoding vector and the power scaling factor respectively, corresponding to the $k$-th user. The total transmit signal will then read as
\begin{equation}\label{eq: transmit signal}
\mathbf x = \sum_{k=1}^K \sqrt{p_k}\mathbf w_{k}s_k.
\end{equation}
Subsequently, when beamforming is employed, \eqref{eq: General input-output} will become
\begin{equation}\label{eq: General input-output2}
y_{k}= \mathbf h^{\dag}_{k}\sqrt{p_k}\mathbf w_{k} s_{k}+ \mathbf h^{\dag}_{k}\sum_{j\neq k} \sqrt{p_j}  \mathbf w_{j} s_{j}+n_k,
\end{equation}
where the first term of the summation refers to the useful signal and the second to the interferences. The column vector \(\mathbf w_{k}\) with \(1 \times N\) dimensions
is the \(k\)-th user's precoding vector, that is the \(k\)-th column of a total  precoding matrix \(\mathbf W = \left[\mathbf w_{1},\mathbf w_{2},\dots, \mathbf w_{k}\right]\).  Subsequently, the SINR at each user is
 \begin{equation}\label{eq: General SINR}
\mathrm{SINR}_{k}= \frac{{p_k}|\mathbf h^{\dag}_{k}\mathbf w_{k}|^2  }{1+\sum_{j\neq k}{p_j}|\mathbf h^{\dag}_{k}\mathbf w_{j}|^2 }.
\end{equation}
When beamforming is employed, determining the optimal precoding vectors and power allocation vectors is tedious in practice. In the following paragraph a very common practical scheme, that will be employed in this paper is presented.

\subsection{Zero Forcing beamforming}
The implementation complexity of the MIMO BC channel capacity achieving dirty paper coding \cite{Weingarten2006a}, led to the development of less complex, yet suboptimal  techniques.  A linear precoding technique with reasonable computational complexity that still achieves full spatial multiplexing and multiuser diversity gains, is ZF precoding\cite{Viswanathan2003,Caire2003,Yoo2006b}. The ability of ZF to fully cancel out multiuser interference makes it useful for the high SNR regime. However, it performs far from optimal in the noise limited regime. Also it can only simultaneously  serve at most equal to the number of transmit antennas,  single antenna users. A common solution for the ZF precoding matrix is the pseudoinverse of the $K\times N$ channel matrix \(\mathbf H = \left[\mathbf h_1 ,\mathbf h_2\dots,  \mathbf h_k \right]^\dag\). Under a total power constraint, channel inversion is the optimal precoder choice in terms of maximum SR and maximum fairness \cite{Wiesel-08}. However,  according to the same authors, when Per Antenna power Constraints (PAC) are assumed, optimization over the parameters of a generalized inverse has to be performed. For simplicity, the present work only considers the simplified channel inversion design. Finally, due to  the symmetrical system, the precoding matrix will read as \(\mathbf W_{}= \mathbf H^{-1}\).

Following the complete mitigation of  interferences,  the $\mathrm{SINR}$ at the $k$-th user will read as \(\mathrm{SINR}_k^{ZF} = {p_k}|\mathbf h^{\dag}_{k}\mathbf w_{k}|^2 \). Subsequently, the problem reduces to a power  allocation optimization process, where the aim is to maximize the total system SR, subject to per antenna power constraints. This problem is formalized as
\begin{eqnarray}\label{prob:opt:throughput}
    \max_{\{p_k\}} && \sum_{k=1}^K\log_2(1+\mathrm{SINR}_{k}^{ZF}),\\
    \mbox{s.t.} && \sum_{k=1}^K p_{k} \mathbf w_{k}^\dag \mathbf Q_{jj}\mathbf w_{k} \leq P_j,\ j=1,\cdots, K. \notag
 \end{eqnarray}
where \(\mathbf Q_{jj}\) is an \(K \times K\) zero matrix, with the \(jj\)-th element equal to one and \(P_j = P_{tot}/K\), for every \(j\),  since we assume that all satellite antennas have identical RF chains. Problem \eqref{prob:opt:throughput}, is a standard convex optimization problem \cite{Yates-07},\cite{Yu2007} that is solved using common convex optimization tools\cite{convex_book}.

\subsection{Dual satellite systems: channel model}\label{sec: channel model}
Extending the previous considerations to a  dual satellite scenario, the resulting  channel needs to be modeled. In this direction, two overlapping clusters of \(N_{1}\) and  \(N_{2}\) spot-beams covering \(K_{1}\) and \(K_{2}\) fixed user  terminals respectively, are considered (Fig. \ref{fig: dual satellite}) . The users, each equipped with a single antenna, are  uniformly distributed over the coverage area.  Despite the fact that in each satellite separately,  a MU MISO BC     is realized, the whole system operates over an interference channel.
   The $k$-th user now has two vector channels, one towards each satellite, denoted using the indices 1 and 2 respectively. The channel vectors \(\mathbf h_{k1}^\dag , \mathbf h_{k2}^\dag\) are the rows of a total channel matrix \(\mathbf H_{tot}\) of \( \left(K_{1} +K_{2} \right) \times \left( N_1+N_2\right) \) dimensions, that models the satellite antenna gains of the two satellites. Assuming that the $k$-th user is allocated  to the first satellite its SINR will read as
\begin{equation}\label{eq: interfering systems SINR}
\hat{\mathrm{SINR} }_{k}= \frac{{p_{1k}}|\mathbf h^{\dag}_{k1}\mathbf w_{k1}|^2 }{1+ \sum_{j=1}^{K_2} p_{2j}|\mathbf h^{\dag}_{k2}\mathbf w_{j2}|^2}.
\end{equation}
In \eqref{eq: interfering systems SINR}, the interferences from the adjacent satellite are clearly noted, while inter-satellite multiuser interferences are completely mitigated by the the precoding.The equivalent relations for the users allocated to the second satellite, are straightforward to deduce, by exchanging indices 1 and 2.

\section{User Selection and Allocation}\label{sec: SIUA}
As proven in \cite{Yoo2005}, user selection can significantly improve the performance of ZF in  an individual system. However, considering the coexistence of two separate transmitters, as is the case in a dual satellite system, partial cooperation, namely coordination, can be employed to solve the problem of high intersatellite interferences. To this end, this contribution proposes an algorithm that selects users and allocates them  to each satellite. Intuitively, the  two basic criteria that need to be considered for this procedure are: a) the maximization of the performance of each satellite separately, and b) the minimization of the interferences between the two sets.


The performance of each satellite separately is optimized by constructing a semi-orthogonal user group from a vast number of users\cite{Yoo2006b}. Extending this result, the creation of two user sets under the semi-orthogonality criterion is straightforward since the channel gain of each user can be projected to the orthogonal complement of the channels of the previously selected users. In each iteration of the algorithm, the user with the maximum projection is allocated to the corresponding set. This simplistic approach has  been considered for comparison reasons.

  The novel proposed algorithm accounts for the effects of  the interferences between the two sets. It should first be mentioned that the exact calculation of the level of interferences in each iteration is not possible since the exact user set is still undetermined. To exactly calculate the interferences, one would need to solve the power optimization problem \eqref{prob:opt:throughput} for all possible combinations of users. Under the assumption of large number of users, this would lead to unaffordable computational complexity. However, based on a basic advantage of ZF beamforming, which is the decoupled nature of the precoder design and the power allocation optimization problems, an approximation of the interferences can be made. In this direction, the precoding vectors of the users selected in the previous iterations can be utilized to provide an indicative measure of the interferences between the user sets. This implies that an equal power allocation is assumed. This assumption becomes asymptotically exact in the high SNR regime, where the powers allocated to each user are approximately equal. Incorporating all the above, a heuristic, iterative Semi-orthogonal Interference aware User Allocation algorithm (SIUA) has been developed and will be presented in the following.

The SIUA algorithm, presented in full detail in Algo. \ref{fig: SIUA}  works as follows. During the initialization procedure, i.e. Step 1,  the strongest user towards each satellite is allocated to the equivalent group. While the two sets are not full, Steps 2 and 3 are executed. In Step 2, for each of the unallocated users, the following metrics are calculated: a) In accordance to \cite{Yoo2006b}, \(\mathbf g_{1k}\) and \(\mathbf g_{2k}\) represent the orthogonal component of each unallocated user's channel to the orthogonal subspace of the already allocated users, for the two sets respectively. In b) \(\mathrm I_{k1}^r \) and \(\mathrm I_{k2}^r\)  are equivalent measures for the interference each user would receive if equal power allocation is employed. It is calculated as the squared norm of  the product  of the users' channel with the power of the transmit signal of the second user set and the channel of each user. Finally, in c) \(\mathrm I_{k1}^i \) and \(I^i_{k2}\)  are approximations of the interference that the allocation of each user can potentially induce to the second set, if this user is allocated in the respective set. It is calculated as the product of the interferences this user induces to every user that belongs to the second set. Since the goal is to find the most orthogonal users that at the same time receive and induce the least possible interferences, the measure to be maximized is the fraction of the orthogonality metric over the product of the interference metrics. At the last stage of each iteration, two maximum fractions \(\mu_{1}\ \text{and}\ \mu_2\) are calculated over the hole user set and compared between them. The user that corresponds to the largest measure among the two, is allocated to the equivalent satellite.

The described heuristic, iterative, optimization algorithm, only
requires full knowledge of the CSI of all users. Consequently,  coordination reduces the amount of data that needs to be exchanged since each GW handles only the data of the users allocated to the corresponding satellite. Moreover, SIUA runs only as many times as the number of transmit antennas and thus compromises a scalable solution that can be extended for larger multibeam systems. Another advantage of this solution is that the power optimization in each satellite, a convex optimization problem that requires some computational complexity, is decoupled from the algorithm execution. Additionally, despite the fact that the solution is heuristic and not optimal, it is considerably less complex since the optimal user allocation would require exhaustive search of all possible combinations of the users.
Finally, SIUA, can be executed in a centralized location or run in parallel at the GWs that share CSI.

\begin{table}
\caption{Link Budget Parameters}
\centering
\begin{tabular}{l|c}
\textbf{Parameter}  & \textbf{Value}  \\\hline

Orbit & GEO \\
Frequency band & Ka (20\,GHz)\\
User link bandwidth & 500\,MHz \\
Number of beams &7\\
Beam diameter & 600 \,Km\\\hline
TWTA RF power @ saturation& $+21$\,dBW\\

Max satellite antenna gain $G_{T}$& $+52$\, dBi\\
Max user antenna Gain \(G_R\) & $+40$\, dBi \\
Free space loss & $-210$\,dB\\ \hline
Signal Power $\mathrm S$ & $-97$\,dBW\\\hline
Receiver noise power $N$ & $-118$\,dBW \\\hline
 SNR $\mathrm {S/N}$ & 21\,dB \\\hline
\end{tabular}
\label{tab: LinkBudgParas}
\end{table}
\section{Simulation results}  \label{sec: Simulation Results}
To evaluate the performance of the proposed algorithm, two satellites each with seven beams where assumed. The low number of beams is only chosen to reduce the simulation time of the convex optimization problem \eqref{prob:opt:throughput}   and has no effect on the algorithm, as discussed in Sec. \ref{sec: SIUA}. It is however inline with the future considerations for the terabit satellites, where each GW is expected to handle between 5 and 8 beams. Additionally, the simulations are performed according to the link budget calculations described in Table \ref{tab: LinkBudgParas}, where it can be noted  that the normal SNR operating point of current satellite systems is 21dB.

\begin{figure}
  \centering
  \includegraphics[width=1\columnwidth]{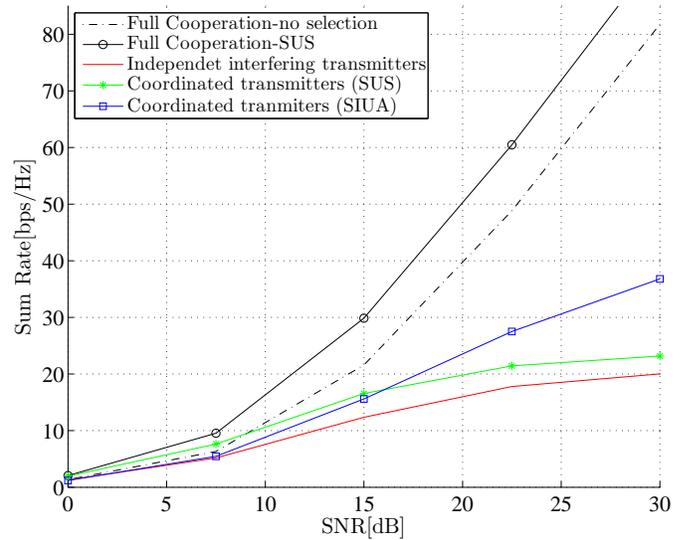}\\
  \caption{Evaluation of SIUA algorithm in terms of system sum rate, by comparison with optimal and interfering systems.}
\label{fig: ICC 1}
\end{figure}

In Fig. \ref{fig: ICC 1}, the results of    Monte Carlo simulations that calculate the performance in terms of system Sum Rate (SR) of the coexisting systems with and without cooperation, are presented. Due to random user positioning,  100 iterations where executed, each with a different user position pattern so that the average performance can be evaluated. The upper bound for the system performance is deduced assuming full cooperation amongst the transmitters. For this case two curves show the SR in bps/Hz: one for the average performance of random user positioning and  one employing the algorithm developed in \cite{Yoo2006b}, namely the Semi-orthogonal  User allocation (SUS) algorithm, which allocates users without regarding the coexistence of the systems. Subsequently,  an average gain of 25\% is noted by employing a simple user selection scheme, instead of randomly selecting users.  Since in the fully cooperative system, interferences are completely mitigated,  SIUA is unnecessary.    

The substantial performance gain from SIUA is proven in  Fig. \ref{fig: ICC 1} for the more realistic case of coordinated systems. In this figure, SIUA is compared to SUS and also to an independent interfering system. From these curves, it is concluded that in the low SNR region, the SUS algorithm performs better, as expected since the noise limited regime is almost interference free. In the SNR region of interest, however, it is clear that the SIUA algorithm, by reducing the level of interferences, provides substantial gains: more than 52\% of improvement in terms of SR, over a non cooperative system and 28\% of improvement over a coordinated system employing simple user selection. It is therefore concluded that the SIUA sacrifices some low SNR performance to provide substantial gains in the SNR region of interest. A simple switching scheme between the two algorithms can provide good performance over the whole SNR region of interest.\

\begin{figure}
  \centering
  \includegraphics[width=1\columnwidth]{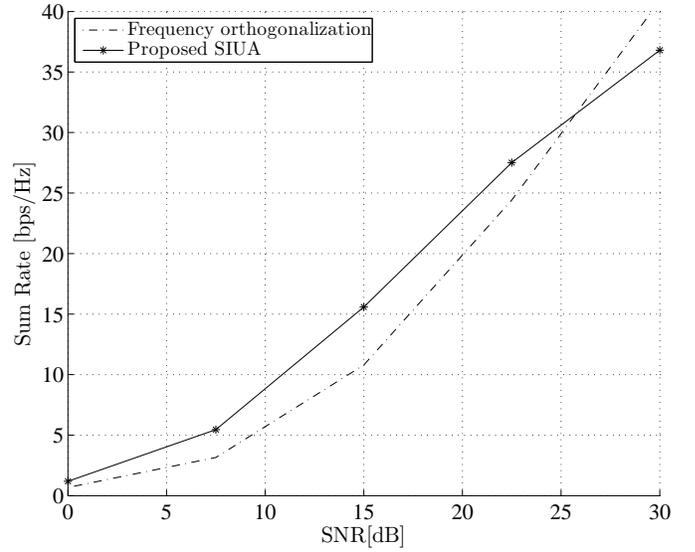}\\
  \caption{Comparison of a coordinated system employing SIUA, with a conventional frequency orthogonalization system.}
\label{fig: ICC2}
\end{figure}

In Fig. \ref{fig: ICC2}, a coordinated system using SIUA, is compared to an ideal non interfering dual satellite system that employs frequency orthogonalization to allow the operability of the two coexisting satellites. This approach models the currently employed techniques of bandwidth splitting. In this plot it is proven that around the SNR area of interest, the proposed algorithm  outperforms the conventional techniques (25\% gain). Therefore, the SIUA comprises a candidate tool for handling ASI. As the SNR increases, the gain decreases as expected, since the conventional system operates under the ideal assumption of zero interferences.

\begin{figure}
  \centering
  \includegraphics[width=0.45\textwidth]{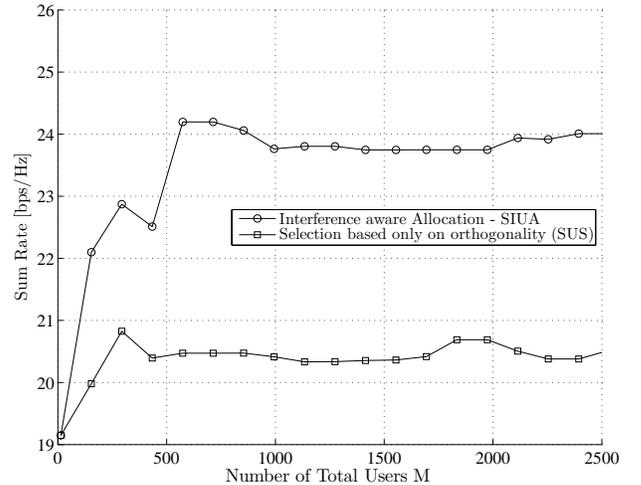}\\
  \caption{Algorithm performance with respect to the number of available for selection users.}
\label{fig: ICC3}
\end{figure}

Finally, in Fig. \ref{fig: ICC3},  the behavior of the algorithm with respect to the size of the user pool is investigated. To this end, the achievable SR for a given value of SNR, i.e. 20dB,  is calculated as the total number  users increases and
also compared to the performance of the SUS algorithm. In this figure it is proven that the algorithm reaches close to its maximum performance for a finite number of total users (600 users) and further increase of the user pool has little effects. From the same figure we note that the rate of convergence of the proposed technique to the saturation point is very similar to the  SUS algorithm.

 \section{Conclusions and future work }\label{sec: conclusions}
 The topic of dual satellite systems is addressed in the present contribution and a suboptimal, simple solution is proposed. A low complexity, heuristic algorithm that minimizes the interferences between the two groups, while maintaining the orthogonality between the users of the same group, allows for the coexistence of two separate multibeam, joint processing, coordinated  satellites. Thereby the overall system spectral efficiency is increased. The only cost of the proposed solution is the CSI exchange between the GWs. According to simulation results, the algorithm achieves 52\% of spectral efficiency improvement over non-cooperative full frequency reuse systems and 25\% improvement over non-cooperative conventional  orthogonalized in the frequency domain systems. Subsequently, the proposed scheme, successfully exploits the spatial orthogonalization of users and allows for systems that are partially cooperative to operate over all the available spectrum.

This fundamental approach can be extended as a useful interference mitigation technique for several other scenarios. Also, the consideration of a third set of users that will be simultaneously served by both satellites to achieve some minimum requirements will be part of future work.  The dual satellite system will also be investigated as a cognitive communications system where priority will be given to primary users.
\section*{Acknowledgment}
This work was   partially supported by the National Research Fund, Luxembourg under the project  ``$CO^{2}SAT:$ Cooperative \& Cognitive Architectures for Satellite Networks' and by the seventh framework program project ``$CORASAT:   $ COgnitive RAdio for SATellite Communications.

\begin{algorithm}\label{fig: SIUA}
\SetAlgoLined
\textbf{SIUA algorithm}\\
\KwOut{ $\mathcal{S}_1 \ \&  \ \mathcal{S}_2$}
\emph{Step 1:}
 $\forall \ k = 1, 2 \dots M $ allocate the strongest channel norm to each satellite: \\
 $\pi_{1(1)} = \arg \max ||{\mathbf h_{k1}}|| $, $\mathbf g_{1(1)} = \mathbf h_{\pi_{1}1} $\\
 $\pi_{2(1)} = \arg \max ||{\mathbf h_{k2}}|| $, $ \ \mathbf g_{2(1)} = \mathbf h_{ \pi_{2}2} $\\
 $\mathcal{S}_1 = \pi_{1(1)},\ \mathcal{S}_2 = \pi_{2(1)}$ \\
 $\mathcal{T}=\{1,\dots M \} - \{ \pi_{1(1)}, \pi_{2(1)} \}\  $set of unprocessed users\\
 $i=1$ iteration counter\\

\While{$\left(|\mathcal{S}_1|<M_1\right) \&\left(|\mathcal{S}_2|<M_2\right)$\\}{
\emph{Step 2:}
\ForAll{ elements of $\mathcal{T}_{(i)}$}{
(a)
$\mathbf g_{1k} = \mathbf h_{1k}\left(\mathbf{I}_K-\sum_{j=1}^{i-1} \frac{\mathbf{g}^\dag_{1(j)} \mathbf{g}_{1(j)}}{||\mathbf{g}_{1(j)||^2}} \right)$ \\
\ \ \ \ $\mathbf g_{2k} = \mathbf h_{k2}\left(\mathbf{I}_K-\sum_{j=1}^{i-1} \frac{\mathbf{g}^\dag_{2(j)} \mathbf{g}_{2(j)}}{||\mathbf{g}_{2(j)||^2}} \right)$ \\
(b) $\mathrm I^{r}_{1k} = \mathbf h_{k2}\left(\mathbf{W}_2\mathbf{W}^\dag_{2}\right)\mathbf h^\dag_{k2}$ \\
\ \ \ \  $\mathrm I^r_{2k} = \mathbf h_{k1}\left(\mathbf{W}_1\mathbf{W}^\dag_{1}\right)\mathbf h^\dag_{k1}$  \\
(c) $\mathrm I^i_{1k} = \prod_{l\epsilon \ \mathbf{t}}^{l\neq k}\left(\mathbf h_{l1}\left(\mathbf{W}_{1k}\mathbf{W}^\dag_{1k}\right)\mathbf h^\dag_{l1}\right)$ \\
\ \ \ \  $\mathrm I^i_{2k} =\prod_{l\epsilon \ \mathbf{t}}^{l\neq k}\left(\mathbf h_{l2}\left(\mathbf{W}_{2k}\mathbf{W}^\dag_{2k}\right)\mathbf h^\dag_{l2}\right)$ \\
where $\mathbf{W}_{n}, n = 1,2$ is the ZF precoding matrix of each satellite
with users allocated from previous iterations and $\mathbf{W}_{nk}, k  \epsilon \ \mathbf{t}$
is the same matrix but with the $k$th user added.
}

\emph{Step 3:} $\mu_{1(i)} = \max \{\mathbf ||g_{1k}||/\left(\mathrm{I}^r_{1k}\cdot\mathrm I^i_{1k}\right) \}$, $\mu_{2(i)} = \max \{\mathbf ||g_{2k}||/\left(\mathrm{I}^r_{2k}\cdot\mathrm I^i_{2k}\right) \}$\\
\eIf{$\mu_{1(i)} \geq  \mu_{2(i)}$ \& $|\mathcal{S}_1|<M_1$}{
$\pi_{(i)} = \arg \mu_{1(i)};$ $\mathcal{S}_1 = \mathcal{S}_1 \cup \{\pi_{(i)}\};$\\
$\mathbf g_{1(i)} = \mathbf h_{\pi_{(i)}};$
}
{
$\pi_{(i)} = \arg \mu_{2(i)};$ $\mathcal{S}_2 = \mathcal{S}_2 \cup \{\pi_{(i)}\};$\\
$\mathbf g_{2(i)} = \mathbf h_{\pi_{(i)}};$
}
$i=i+1;$ \\$\mathcal{T}_{(i)} = \mathcal{T}_{(i-1)}- \{\pi_{(i-1)}\};$
}
\caption{Semiorthogonal Interference aware User Allocation algorithm (SIUA)}
\end{algorithm}

\bibliographystyle{IEEEtran}
\bibliography{refs/IEEEabrv,refs/conferences,refs/journals,refs/books,refs/references,refs/csi,refs/thesis}

\end{document}